\documentstyle[aps,prl,epsf]{revtex}
\bibstyle{unsrt}

\begin{document}
\draft

\twocolumn[\hsize\textwidth\columnwidth\hsize\csname 
@twocolumnfalse\endcsname

\preprint{HEP-96-13}

\title{Real Spectra in Non-Hermitian Hamiltonians Having ${\cal PT}$ Symmetry}

\author{Carl M. Bender$^1$ and Stefan Boettcher$^{2,3}$}
\address{${}^1$Department of Physics, Washington University, St. Louis, MO
63130, USA}
\address{${}^2$Center for Nonlinear Studies, Los Alamos National Laboratory,
Los Alamos, NM 87545, USA}
\address{${}^3$ CTSPS, Clark Atlanta University, Atlanta, GA 30314, USA}

\date{\today}
\maketitle

\begin{abstract}
The condition of self-adjointness ensures that the eigenvalues of a Hamiltonian
are real and bounded below. Replacing this condition by the weaker condition of
${\cal PT}$ symmetry, one obtains new infinite classes of complex Hamiltonians
whose spectra are also real and positive. These ${\cal PT}$ symmetric theories
may be viewed as analytic continuations of conventional theories from real to
complex phase space. This paper describes the unusual classical and quantum
properties of these theories.
\end{abstract}
]

Several years ago, D.~Bessis conjectured on the basis of numerical studies that
the spectrum of the Hamiltonian $H=p^2+x^2+ix^3$ is {\em real and positive}
\cite{BESSIS}. To date there is no rigorous proof of this conjecture. We claim
that the reality of the spectrum of $H$ is due to ${\cal PT}$ symmetry. Note
that $H$ is invariant {\sl neither} under parity ${\cal P}$, whose effect is to
make spatial reflections, $p\to-p$ and $x\to-x$, {\sl nor} under time reversal
${\cal T}$, which replaces $p\to-p$, $x\to x$, and $i\to-i$. However,
${\cal PT}$ symmetry is crucial. For example, the Hamiltonian $p^2+ix^3+ix$ has
${\cal PT}$ symmetry and our numerical studies indicate that its entire spectrum
is positive definite; the Hamiltonian $p^2+ix^3+x$ is not ${\cal PT}$-symmetric,
and the entire spectrum is complex.

The connection between ${\cal PT}$ symmetry and positivity of spectra is simply
illustrated by the harmonic oscillator $H=p^2+x^2$, whose energy levels are $E_n
=2n+1$. Adding $ix$ to $H$ does not break ${\cal PT}$ symmetry, and the spectrum
remains positive definite: $E_n=2n+{5\over4}$. Adding $-x$ also does not break
${\cal PT}$ symmetry if we define ${\cal P}$ as reflection about $x={1\over2}$,
$x\to1-x$, and again the spectrum remains positive definite: $E_n=2n+{3\over4}$.
By contrast, adding $ix-x$ {\em does} break ${\cal PT}$ symmetry, and the
spectrum is now complex: $E_n=2n+1+{1\over2}i$.

The Hamiltonian studied by Bessis is just one example of a huge and remarkable
class of non-Hermitian Hamiltonians whose energy levels are real and positive.
The purpose of this Letter is to understand the fundamental properties of such a
theory by examining the class of quantum-mechanical Hamiltonians
\begin{eqnarray}
H=p^2+m^2 x^2-(ix)^N\quad(N~{\rm real}).
\label{e1}
\end{eqnarray}
As a function of $N$ and mass $m^2$ we find various phases with transition
points at which entirely real spectra begin to develop complex eigenvalues.

There are many applications of non-Hermitian ${\cal PT}$-invariant Hamiltonians
in physics. Hamiltonians rendered non-Hermitian by an imaginary external field
have been introduced recently to study delocalization transitions in condensed
matter systems such as vortex flux-line depinning in type-II superconductors
\cite{Hatano+Nelson}, or even to study population biology \cite{Nelson+Shnerb}.
Here, initially real eigenvalues bifurcate into the complex plane due to the
increasing external field, indicating the unbinding of vortices or the growth of
populations. We believe that one can also induce dynamic delocalization by
tuning a physical parameter (here $N$) in a self-interacting theory.

Furthermore, it was found that quantum field theories analogous to the
quantum-mechanical theory in Eq.~(\ref{e1}) have astonishing properties. The
Lagrangian $L=(\nabla\phi)^2+m^2 \phi^2-g(i\phi)^N$ ($N$ real) possesses
${\cal PT}$ invariance, the fundamental symmetry of local self-interacting
scalar quantum field theory \cite{PCT}. Although this theory has a non-Hermitian
Hamiltonian, the spectrum of the theory appears to be positive definite. Also,
$L$ is explicitly not parity invariant, so the expectation value of the field
$\langle\phi\rangle$ is nonzero, even when $N=4$ \cite{BM}. Thus, one can
calculate directly (using the Schwinger-Dyson equations, for example
\cite{MILTON}) the (real positive) Higgs mass in a renormalizable theory such as
$-g\phi^4$ or $ig\phi^3$ in which symmetry breaking occurs naturally (without
introducing a symmetry-breaking parameter).

Replacing conventional $g\phi^4$ or $g\phi^3$ theories by $-g\phi^4$ or $ig
\phi^3$ theories has the effect of reversing signs in the beta function. Thus,
theories that are not asymptotically free become asymptotically free and
theories that lack stable critical points develop such points. For example,
${\cal PT}$-symmetric massless electrodynamics has a nontrivial stable critical
value of the fine-structure constant $\alpha$ \cite{ALPHA}.

Supersymmetric non-Hermitian, ${\cal PT}$-invariant Lagrangians have been
examined \cite{BMsuper}. It is found that the breaking of parity symmetry does
not induce a breaking of the apparently robust global supersymmetry. The
strong-coupling limit of non-Hermitian ${\cal PT}$-symmetric quantum field
theories has been investigated \cite{BBJM}; the correlated limit in which the
bare coupling constants $g$ and $-m^2$ both tend to infinity with the
renormalized mass $M$ held fixed and finite, is dominated by solitons. (In
parity-symmetric theories the corresponding limit, called the Ising limit, is
dominated by instantons.)

\begin{figure}
\epsfxsize=2.2truein
\hskip 0.15truein\epsffile{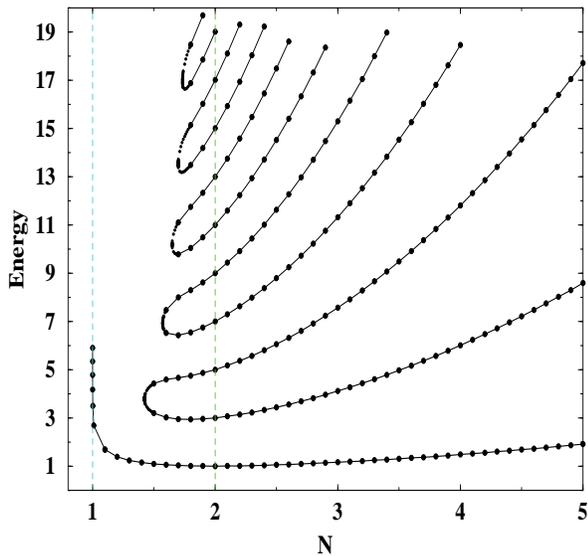}
\caption{
\narrowtext
Energy levels of the Hamiltonian $H=p^2-(ix)^N$ as a function of the parameter
$N$. There are three regions: When $N\geq2$ the spectrum is real and positive.
The lower bound of this region, $N=2$, corresponds to the harmonic oscillator,
whose energy levels are $E_n=2n+1$. When $1<N<2$, there are a finite number of
real positive eigenvalues and an infinite number of complex conjugate pairs of
eigenvalues. As $N$ decreases from $2$ to $1$, the number of real eigenvalues
decreases; when $N\leq1.42207$, the only real eigenvalue is the ground-state
energy. As $N$ approaches $1^+$, the ground-state energy diverges. For $N\leq1$
there are no real eigenvalues.}
\label{fig1}
\end{figure}

To elucidate the origin of such novel features we examine the elementary
Hamiltonian (\ref{e1}) using extensive numerical and asymptotic studies. As
shown in Fig.~\ref{fig1}, when $m=0$ the spectrum of $H$ exhibits three distinct
behaviors as a function of $N$. When $N\geq2$, the spectrum is infinite,
discrete, and entirely real and positive. (This region includes the case $N=4$
for which $H=p^2-x^4$; the spectrum of this Hamiltonian is positive and discrete
and $\langle x\rangle\neq0$ in the ground state because $H$ breaks parity
symmetry!) At the lower bound $N=2$ of this region lies the harmonic oscillator.
A phase transition occurs at $N=2$; when $1<N<2$, there are only a {\em finite}
number of real positive eigenvalues and an infinite number of complex conjugate
pairs of eigenvalues. In this region ${\cal PT}$ symmetry is {\it spontaneously
broken} \cite{BBM}. As $N$ decreases from $2$ to $1$, adjacent energy levels
merge into complex conjugate pairs beginning at the high end of the spectrum;
ultimately, the only remaining real eigenvalue is the ground-state energy, which
diverges as $N\to1^+$ \cite{SH}. When $N\leq1$, there are no real eigenvalues.
The massive case $m\neq0$ is even more elaborate; there is a phase transition at
$N=1$ in addition to that at $N=2$.

The Schr\"odinger eigenvalue differential equation corresponding to the
Hamiltonian (\ref{e1}) with $m=0$ is 
\begin{eqnarray}
-\psi''(x)-(ix)^N\psi(x)=E\psi(x).
\label{e2}
\end{eqnarray}
Ordinarily, the boundary conditions that give quantized energy levels $E$ are
that $\psi(x)\to0$ as $|x|\to\infty$ on the real axis; this condition suffices
when $1<N<4$. However, for arbitrary real $N$ we must continue the eigenvalue
problem for (\ref{e2}) into the complex-$x$ plane. Thus, we replace the real-$x$
axis by a contour in the complex plane along which the differential equation
holds and we impose the boundary conditions that lead to quantization at the
endpoints of this contour. (Eigenvalue problems on complex contours are
discussed in Ref.~\cite{ROT}.)

\begin{figure}
\epsfxsize=2.2truein
\hskip 0.15truein\epsffile{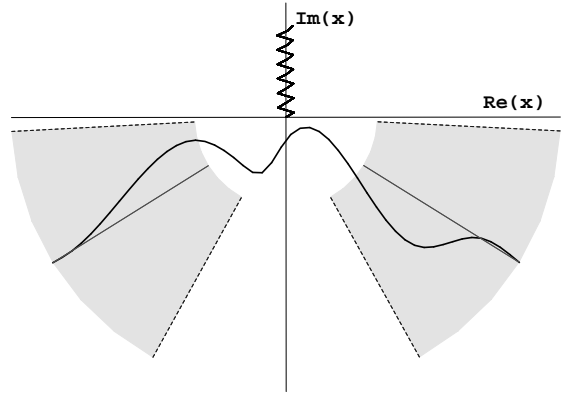}
\caption{
\narrowtext
Wedges in the complex-$x$ plane containing the contour on which the eigenvalue
problem for the differential equation (\protect\ref{e2}) for $N=4.2$ is posed.
In these wedges $\psi(x)$ vanishes exponentially as $|x|\to\infty$. The wedges
are bounded by {\em Stokes lines} of the differential equation. The center of
the wedge, where $\psi(x)$ vanishes most rapidly, is an anti-Stokes line.}
\label{fig2}
\end{figure}

The regions in the cut complex-$x$ plane in which $\psi(x)$ vanishes
exponentially as $|x|\to\infty$ are {\em wedges} (see Fig.~\ref{fig2}); these
wedges are bounded by the {\em Stokes lines} of the differential equation
\cite{BO}. The center of the wedge, where $\psi(x)$ vanishes most rapidly, is
called an {\em anti-Stokes line.}

There are many wedges in which $\psi(x)\to0$ as $|x|\to\infty$. Thus, there are
many eigenvalue problems associated with a given differential equation
\cite{ROT}. However, we choose to continue the eigenvalue equation (\ref{e2})
away from the conventional harmonic oscillator problem at $N=2$. The wave
function for $N=2$ vanishes in wedges of angular opening ${1\over2}\pi$ centered
about the negative- and positive-real $x$ axes. For arbitrary $N$ the
anti-Stokes lines at the centers of the left and right wedges lie at the angles
\begin{eqnarray}
\theta_{\rm left}=-\pi+{N-2\over N+2}~{\pi\over2}\quad{\rm and}\quad
\theta_{\rm right}=-{N-2\over N+2}~{\pi\over2}.
\label{e3}
\end{eqnarray}
The opening angle of these wedges is $\Delta=2\pi/(N+2)$. The differential
equation (\ref{e2}) may be integrated on any path in the complex-$x$ plane so
long as the ends of the path approach complex infinity inside the left wedge and
the right wedge \cite{QFT}. Note that these wedges contain the real-$x$ axis
when $1<N<4$.

As $N$ increases from $2$, the left and right wedges rotate downward into the
complex-$x$ plane and become thinner. At $N=\infty$, the differential equation
contour runs up and down the negative imaginary axis and thus there is no
eigenvalue problem at all. Indeed, Fig.~\ref{fig1} shows that the eigenvalues
all diverge as $N\to\infty$. As $N$ decreases below $2$ the wedges become wider
and rotate into the upper-half $x$ plane. At $N=1$ the angular opening of the
wedges is ${2\over3}\pi$ and the wedges are centered at ${5\over6}\pi$ and
${1\over6}\pi$. Thus, the wedges become contiguous at the positive-imaginary $x$
axis, and the differential equation contour can be pushed off to infinity.
Consequently, there is no eigenvalue problem when $N=1$ and, as we would expect,
the ground-state energy diverges as $N\to1^+$ (see Fig.~\ref{fig1}).

To ensure the numerical accuracy of the eigenvalues in Fig.~\ref{fig1}, we have
solved the differential equation (\ref{e2}) using two independent procedures.
The most accurate and direct method is to convert the complex differential
equation to a system of coupled, real, second-order equations which we solve
using the Runge-Kutta method; the convergence is most rapid when we integrate
along anti-Stokes lines. We then patch the two solutions together at the origin.
We have verified those results by diagonalizing a truncated matrix
representation of the Hamiltonian in Eq.~(\ref{e1}) in harmonic oscillator
basis functions.
\medskip

\noindent
{\em Semiclassical analysis:} Several features of Fig.~\ref{fig1} can be
verified analytically. When $N\geq2$, WKB gives an excellent approximation to
the spectrum. The novelty of this WKB calculation is that it must be performed
in the complex plane. The turning points $x_{\pm}$ are those roots of
$E+(ix)^N=0$ that {\sl analytically continue} off the real axis as $N$ moves
away from $N=2$ (the harmonic oscillator):
\begin{eqnarray}
x_-=E^{1/N}e^{i\pi(3/2-1/N)},\quad x_+=E^{1/N}e^{-i\pi(1/2-1/N)}.
\label{e5}
\end{eqnarray}
These turning points lie in the lower-half (upper-half) $x$ plane in
Fig.~\ref{fig2} when $N>2$ ($N<2$).

The leading-order WKB phase-integral quantization condition is
$(n+1/2)\pi=\int_{x_-}^{x_+}dx\,\sqrt{E+(ix)^N}$. It is crucial that
this integral follow a path along which the {\em integral is real.} When $N>2$,
this path lies entirely in the lower-half $x$ plane and when $N=2$ the path lies
on the real axis. But, when $N<2$ the path is in the upper-half $x$ plane; it
crosses the cut on the positive-imaginary 
\begin{table}
\caption[t1]{Comparison of the exact eigenvalues (obtained with Runge-Kutta) and
the WKB result in (\ref{e7}).}
\begin{tabular}{llddlldd}
$N$ & $n$ & $E_{\rm exact}$ & $E_{\rm WKB}$ & $N$ & $n$ & $E_{\rm exact}$ & $E_
{\rm WKB}$ \\ \tableline
3 & 0 & 1.1562& 1.0942& 4 & 0 &  1.4771 & 1.3765 \\
&1 & 4.1092& 4.0894&   & 1 &  6.0033 &  5.9558 \\
&2 & 7.5621& 7.5489&  & 2 & 11.8023 & 11.7689 \\
&3 & 11.3143& 11.3042&  & 3 &  18.4590 &  18.4321 \\
&4 & 15.2916& 15.2832&  &  &  & \\
\end{tabular}
\label{table1}
\end{table}
\noindent
axis and thus is {\em not a continuous
path joining the turning points.} Hence, WKB fails when $N<2$.

When $N\geq2$, we deform the phase-integral contour so that it follows the rays
from $x_-$ to $0$ and from $0$ to $x_+$: $(n+1/2)\pi=2\sin(\pi/N)E^{1/N+1/2}
\int_0^1 ds\,\sqrt{1-s^N}$. We then solve for $E_n$:
\begin{eqnarray}
E_n\sim\left[{\Gamma(3/2+1/N)\sqrt{\pi}(n+1/2)\over\sin(\pi/N)\Gamma(1+1/N)}
\right]^{2N\over N+2}\quad(n\to\infty).
\label{e7}
\end{eqnarray}
We perform a higher-order WKB calculation by replacing the phase integral by a
{\em closed contour} that encircles the path in Fig.~\ref{fig2} (see
Ref.~\cite{BO,BBM}). See Table I.

It is interesting that the spectrum of the $|x|^N$ potential is like that of the
$-(ix)^N$ potential. The leading-order WKB quantization condition (accurate for
$N>0$) is like Eq.~(\ref{e7}) except that $\sin(\pi/N)$ is absent. However, as
$N\to\infty$, the spectrum of $|x|^N$ approaches that of the square-well
potential [$E_n=(n+1)^2\pi^2/4$], while the energies of the $-(ix)^N$ potential
diverge (see Fig.~1).
\medskip

\noindent
{\em Asymptotic study of the ground-state energy near $N=1$}: When $N=1$, the
differential equation (\ref{e2}) can be solved exactly in terms of Airy
functions. The anti-Stokes lines at $N=1$ lie at $30^\circ$ and at $150^\circ$.
We find the solution that vanishes exponentially along each of these rays and
then rotate back to the real-$x$ axis to obtain
\begin{eqnarray}
\psi_{\rm left,\,right}(x)=C_{1,\,2}\,{\rm Ai}(\mp xe^{\pm i\pi/6}
+Ee^{\pm 2i\pi/3}).
\label{e8}
\end{eqnarray}
We must patch these solutions together at $x=0$ according to the patching
condition $\left.{d\over dx}|\psi(x)|^2\right|_{x=0}=0$. But for real $E$, the
Wronskian identity for the Airy function is
\begin{eqnarray}
{d\over dx}|{\rm Ai}(xe^{-i\pi/6}+Ee^{-2i\pi/3})|^2\Bigm|_{x=0}=-{1\over2\pi}
\label{e9}
\end{eqnarray}
instead of $0$. Hence, there is no real eigenvalue.

Next, we perform an asymptotic analysis for $N=1+\epsilon$, $-\psi''(x)-(ix)^{1+
\epsilon}\psi(x)=E\psi(x)$, and take $\psi(x)=y_0(x)+\epsilon y_1(x)+{\rm O}
(\epsilon^2)$ as $\epsilon\to0+$. We assume that $E\to\infty$ as
$\epsilon\to0+$, let $C_2=1$ in Eq.~(\ref{e8}), and obtain
\begin{eqnarray}
y_0(0)={\rm Ai}(Ee^{-2i\pi/3})\sim e^{i\pi/6}E^{-1/4}e^{{2\over3}E^{3/2}}/
(2\sqrt{\pi}).
\label{e10}
\end{eqnarray}
We set $y_1(x)=Q(x)y_0(x)$ in the inhomogeneous equation
$-y_1''(x)-ixy_1(x)-Ey_1(x)=ix\ln(ix)y_0(x)$ and get
\begin{eqnarray}
Q'(0)={i\over y_0^2(0)}\int_0^{\infty}dx\,x\,\ln(ix)y_0^2(x).
\label{e11}
\end{eqnarray}

Choosing $Q(0)=0$, we find that the patching condition at $x=0$ gives
$1=2\pi\epsilon\left|y_0(0)\right|^2[Q'(0)+{Q^*}'(0)]$, where we have used the
zeroth-order result in Eq.~(\ref{e9}). Using Eqs.~(\ref{e10}) and (\ref{e11})
this equation becomes
\begin{eqnarray}
1={\epsilon\over \sqrt{E}}e^{{4\over3}E^{3/2}}{\rm Re}\,\left[{i\over y_0^2(0)}
\int_0^{\infty}dx\,x\,\ln(ix)y_0^2(x)\right].
\label{e12}
\end{eqnarray}
Since $y_0(x)$ decays rapidly as $x$ increases, the integral in Eq.~(\ref{e12})
is dominated by contributions near $0$. Asymptotic analysis of this integral
gives an implicit equation for $E$ as a function of $\epsilon$ (see Table II):
\begin{eqnarray}
1\sim\epsilon e^{{4\over3}E^{3/2}}E^{-3/2}[\sqrt{3}\ln(2\sqrt{E})
+\pi-(1-\gamma)\sqrt{3}]/8.
\label{e13}
\end{eqnarray}

\noindent
{\em Behavior near $N=2$}: The most interesting aspect of Fig.~\ref{fig1} is the
transition that occurs at $N=2$. To describe quantitatively the merging of
eigenvalues that begins when $N<2$ we let $N=2-\epsilon$ and study the
asymptotic behavior as $\epsilon\to0+$. (A Hermitian perturbation causes
adjacent energy levels to repel, but in this case the non-Hermitian perturbation
of the harmonic oscillator $(ix)^{2-\epsilon}\sim x^2-\epsilon x^2[\ln(|x|+{1
\over2}i\pi\,{\rm sgn}(x)]$ causes the levels to merge.) A complete description
of this asymptotic study is given elsewhere \cite{BBM}.

The onset of eigenvalue merging is a phase transition that occurs even at the
{\sl classical} level. Consider the classical equations of motion for a particle
of energy $E$ subject to the complex forces described by the Hamiltonian
(\ref{e1}). For $m=0$ the trajectory $x(t)$ of the particle obeys $\pm dx[E+
(ix)^N]^{-1/2}=2dt$. While $E$ and $dt$ are real, $x(t)$ is a path in the
complex plane in Fig.~\ref{fig2}; this path terminates at the classical turning
points $x_\pm$ in (\ref{e5}).

When $N\geq2$, the trajectory is an arc joining $x_\pm$ in the lower complex
plane. The motion is {\sl periodic}; we have a complex pendulum whose (real)
period $T$ is
\begin{eqnarray}
T=2E^{2-N\over2N}\cos\left[{(N-2)\pi\over2N}\right]{\Gamma(1+1/N)\sqrt{\pi}\over
\Gamma(1/2+1/N)}.
\label{e14}
\end{eqnarray}

At $N=2$ there is a global change. For $N<2$ a path starting at one turning
point, say $x_+$, moves toward but {\sl misses} the turning point $x_-$. This
path spirals outward crossing from sheet to sheet on the Riemann surface, and
eventually veers off to infinity asymptotic to the angle ${N\over2-N}\pi$.
Hence, the period abruptly becomes infinite. The total angular rotation of the
spiral is finite for all $N\neq2$ and as $N\to2^+$, but becomes infinite as
$N\to2^-$. The path passes many turning points as it spirals anticlockwise from
$x_+$. [The $n$th turning point lies at the angle ${4n-N+2\over2N}\pi$ ($x_+$
corresponds to $n=0$).] As $N$ approaches $2$ from below, when the classical
trajectory passes a new 
\begin{table}
\caption[t2]{Comparison of the exact ground-state energy $E$ near $N=1$ and the
asymptotic results in Eq.~(\protect\ref{e13}). The explicit dependence of $E$ on
$\epsilon$ is roughly $E\propto(-\ln\epsilon)^{2/3}$.}
\begin{tabular}{ldd}
$\epsilon=N-1$  &  $E_{\rm exact}$ & Eq.~(\ref{e13}) \\ \tableline
0.1 & 1.6837& 2.0955\\
0.01 &2.6797&2.9624\\
0.001 &   3.4947&3.6723\\
0.0001 &   4.1753&4.3013\\
0.00001 &   4.7798&4.8776\\
0.000001 &   5.3383&5.4158\\
0.0000001 &   5.8943&5.9244\\
\end{tabular}
\label{table2}
\end{table}
\noindent
turning point, there corresponds an additional merging
of the quantum energy levels as shown in Fig.~\ref{fig1}). This correspondence
becomes exact in the limit $N\to2^-$ and is a manifestation of Ehrenfest's
theorem.

\begin{figure}
\epsfxsize=2.2truein
\hskip 0.15truein\epsffile{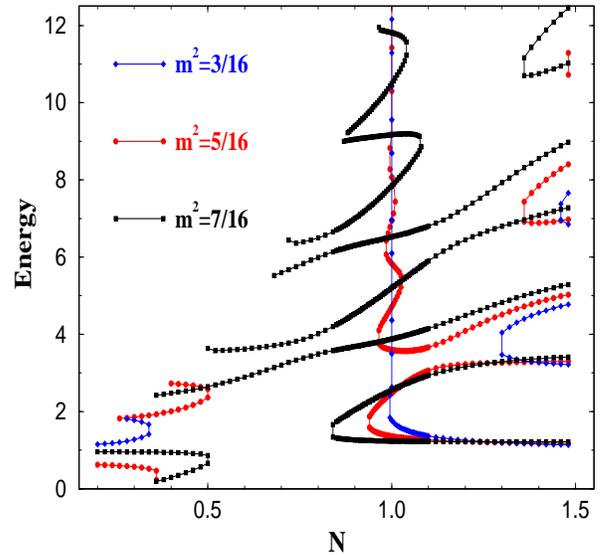}
\caption{
\narrowtext
The $m\neq0$ analog of Fig.~1. Note that transitions occur at $N=2$ and $N=1$.}
\label{fig3}
\end{figure}

\noindent
{\em Massive case}: The $m\neq0$ analog of Fig.~\ref{fig1} exhibits a new
transition at $N=1$ (see Fig.~\ref{fig3}). As $N$ approaches $1$ from above, the
energy levels reemerge from the complex plane in pairs and at $N=1$ the spectrum
is again entirely real and positive. Below $N=1$ the energies once again
disappear in pairs, now including the ground state. As $N\to0$ the infinite real
spectrum reappears again. The massive case is discussed further in
Ref.~\cite{BBM}.

We thank D.~Bessis, H.~Jones, P.~Meisinger, A.~Wightman, and Y.~Zarmi for
illuminating conversations. CMB thanks the Center for Nonlinear Studies, Los
Alamos National Laboratory and STB thanks the Physics Department at Washington
University for its hospitality. This work was supported by the U.S.~Department
of Energy.


\begin{references}
\bibitem{BESSIS} D.~Bessis, private discussion. This problem originated from
discussions between Bessis and J.~Zinn-Justin, who was studying Lee-Yang
singularities using renormalization group methods. An $i\phi^3$ field theory
arises if one translates the field in a $\phi^4$ theory by an imaginary term.
\end{references}
\end{document}